\documentclass[aps,twocolumn,prd,amsmath,nofootinbib,preprintnumbers,superscriptaddress]{revtex4-1}
\usepackage{epsfig,slashed,multirow}

\newcommand{\amc}{{\sc MadGraph5\textunderscore}a{\sc MC@NLO}}
\newcommand{\fr}{{\sc Feyn\-Rules}}
\newcommand{\ma}{{\sc MadAnalysis 5}}
\newcommand{\nloct}{{\sc NloCT}}
\newcommand{\sss}{\scriptscriptstyle}

\def\be{\begin{equation*}}
\def\ee{\end{equation*}}
\def\bsp#1\esp{\begin{split}#1\end{split}} 

\begin{document}

\preprint{CERN-PH-TH/2014-217, DCPT/14/190, IPPP/14/95, LPSC/14/283, MCNET-14-34}

\title{Automated next-to-leading order predictions for new physics at the LHC:\\
   The case of colored scalar pair production}

\author{C\'eline Degrande}
\affiliation{
  Institute for Particle Physics Phenomenology,
  Department of Physics Durham University, Durham DH1 3LE,
  United Kingdom}
\author{Benjamin Fuks}
\affiliation{CERN, PH-TH, CH-1211 Geneva 23, Switzerland}
\affiliation{Institut Pluridisciplinaire Hubert Curien/D\'epartement
  Recherches Subatomiques, Universit\'e de Strasbourg/CNRS-IN2P3,
  23 Rue du Loess, F-67037 Strasbourg, France}
\author{Valentin Hirschi}
\affiliation{SLAC, National Accelerator Laboratory,
  2575 Sand Hill Road, Menlo Park, CA 94025-7090, USA}
\author{Josselin Proudom}
\affiliation{Laboratoire de Physique Subatomique et de Cosmologie,
  Universit\'e Grenoble-Alpes, CNRS/IN2P3, 53 Avenue des Martyrs, F-38026
  Grenoble Cedex, France}
\author{Hua-Sheng Shao}
\affiliation{CERN, PH-TH, CH-1211 Geneva 23, Switzerland}

\begin{abstract}
  We present for the first time the full automation of collider predictions
  matched with parton showers at the next-to-leading accuracy in QCD within
  non-trivial extensions of the Standard Model. The sole inputs required from
  the user are the model Lagrangian and the process of interest. As an
  application of the above, we explore scenarios beyond the Standard Model where
  new colored scalar particles can be pair produced in hadron collisions. Using
  simplified models to describe the new field interactions with the Standard
  Model, we present precision predictions for the LHC within the \amc\ framework.
 \end{abstract}

\maketitle

\section{Introduction}
Motivated by the conceptual issues accompanying the Standard Model,
many new physics theories have been developed over the last decades. Most of
them exhibit an extended colored sector and related new phenomena are expected
to be observable at high-energy hadron colliders such as the LHC.
In particular, effects induced by hypothetical colored scalar particles have
received special attention from both the ATLAS and CMS collaborations. Many LHC
analyses are indeed seeking for the scalar partners of the Standard Model
quarks (the squarks) and gluons (the sgluons) that are predicted, for
instance, in minimal~\cite{Nilles:1983ge,Haber:1984rc} and
non-minimal~\cite{Salam:1974xa,Fayet:1975yi} supersymmetric or in
vector-like confining theories~\cite{Kilic:2009mi}.

In this context, it is clear that an approach to precision predictions that is
fully general in any considered theory is highly desirable, and the \amc\
framework~\cite{Alwall:2014hca} is in a prime position to provide it. Its
structure for tackling leading-order (LO) computations has indeed already
proved to be very efficient at satisfying the needs of both the theoretical and
experimental high energy physics communities. Generalizing this flexibility to
the next-to-leading order (NLO) case is however not straightforward, essentially
because of the necessity of specifying model-dependent counterterms, including
those arising from the renormalization of the Lagrangian. Recent
developments~\cite{Degrande:2014vpa} in the \fr\ package~\cite{Alloul:2013bka}
have allowed to overcome this main obstacle and paved the way to the full
automation of NLO QCD predictions matched to parton showers for generic
theories.

We describe the details of this implementation by working through two specific cases
and revisit some LHC phenomenology associated with stops and sgluons in
the context of simplified models of new physics~\cite{Alwall:2008ag,%
Alves:2011wf}. Employing state-of-the-art simulation techniques, we match
matrix elements to the NLO in QCD to parton showers and present
precision predictions for several kinematical observables after considering both
the production and the decay of the new particles. In more detail, we make use
of \fr\ to implement all possible couplings of
the new fields to quarks and gluons and employ the \nloct\
program~\cite{Degrande:2014vpa} to generate a UFO module~\cite{Degrande:2011ua}
containing, in addition to tree-level model information, the ultraviolet
and $R_2$ counterterms necessary whenever the loop integral numerators are computed
in four dimensions, as in {\sc MadLoop}~\cite{Hirschi:2011pa} that uses
the Ossola-Papadopoulos-Pittau (OPP) reduction
formalism~\cite{Ossola:2006us}. This UFO library is then linked to the
\amc\ framework which is used, for the first time, for
predictions in the context of new physics models featuring an
extended colored sector. We focus on the pair production of the new states
at NLO in QCD. Their decay is then taken into account separately,
at the leading order and with the spin information retained, by means of
the {\sc MadSpin}~\cite{Artoisenet:2012st} and
{\sc MadWidth}~\cite{Alwall:2014bza} programs.

In the rest of this paper, we first define simplified models
describing stop and sgluon dynamics and detail the
renormalization of the effective Lagrangians and the validation of the UFO
models generated by \nloct. Our results follow and consist of total rates
and differential distributions illustrating some kinematical properties of the
produced new states and their decay products.

\section{Benchmark scenarios for stop hadroproduction}
Following a simplified model approach, we extend the Standard Model by a complex
scalar field $\sigma_3$ (a stop) of mass $m_3$. This field lies in the
fundamental representation of $SU(3)_c$, so that its strong interactions are
standard and embedded into $SU(3)_c$-covariant derivatives. We
enable the stop to decay via a coupling to a single top quark and a
gauge-singlet Majorana fermion $\chi$ of mass $m_\chi$ that can be identified
with a bino in complete supersymmetric models. Finally, despite of being
allowed by gauge
invariance, the single stop couplings to down-type quarks, as predicted in
$R$-parity violating supersymmetry, are ignored for simplicity.
We model all considered interactions by the Lagrangian
\be\bsp
  {\cal L}_3 = &\
    D_\mu \sigma_3^\dag D^\mu \sigma_3 - m_3^2 \sigma_3^\dag \sigma_3
    + \frac{i}{2}\bar\chi\slashed{\partial}\chi-\frac12m_\chi\bar\chi\chi\\
  &\quad
    + \Big[ \sigma_3 \bar t \big(\tilde g_L P_L+\tilde g_R P_R \big) \chi
    + {\rm h.c.} \Big]\ ,
\esp\ee
where we denote the strengths of the stop
couplings to the fermion $\chi$ by $\tilde g$ and $P_{L,R}$ are the
left- and right-handed chirality projectors.

Aiming to precision predictions at the NLO accuracy, a renormalization
procedure is required in order to absorb all ultraviolet divergences yielded
by virtual loop-diagrams. This is achieved through
counterterms that are derived from the tree-level Lagrangian by
replacing all bare fields (generically denoted by $\Psi$) and parameters
(generically denoted by $A$) by
\be\bsp
  \Psi \to Z_\Psi^{1/2}\Psi \approx
    \big[1 + \frac12 \delta Z_\Psi\big] \Psi
  \quad\text{and}\quad
  A \to A + \delta Z_A \ ,
\esp\ee
where the renormalization constants $\delta Z$ are restricted in our case to
QCD contributions at the first order in the strong coupling $\alpha_s$.
Like in usual supersymmetric setups, the $\tilde g$ couplings are of a non-QCD
nature so that our simplified model does not feature new strong interactions
involving quarks. The wave-function renormalization constant of the latter is
therefore unchanged with respect to the Standard Model, contrary to the gluon
one that must appropriately compensate stop-induced contributions.
Adopting the on-shell renormalization scheme, the
gluon and stop wave-function ($\delta Z_g$ and $\delta Z_{\sigma_3}$) and mass
($\delta m_3^2$) renormalization constants read
\be\bsp
  \delta Z_g =&\ \delta Z_g^{(SM)} - \frac{g_s^2}{96 \pi^2}
    \bigg[\frac{1}{\bar\epsilon} - \log\frac{m_3^2}{\mu_R^2}\bigg] \ ,\\
  \delta Z_{\sigma_3} =&\ 0
  \quad\text{and}\quad
  \delta m_3^2 = -\frac{g_s^2 m_3^2}{12\pi^2}
      \Big[\frac{3}{\bar\epsilon} +  7 - 3\log\frac{m_3^2}{\mu_R^2} \Big] \ ,
\esp\ee
where $\delta Z_g^{(SM)}$ collects the Standard Model components
of $\delta Z_g$. Moreover,
we denote the renormalization scale by $\mu_R$ and following standard
conventions, the ultraviolet divergent parts of the renormalization constants
are written in terms of the quantity $1/\bar\epsilon = 1/\epsilon - \gamma_E +
\log4\pi$ where $\gamma_E$ is the Euler-Mascheroni constant and $\epsilon$ is
linked to the number of space-time dimensions $D=4-2\epsilon$.

The renormalization of the strong coupling is achieved by subtracting, at
zero-momentum transfer, all heavy particle contributions from the gluon
self-energy. This ensures that the running of $\alpha_s$ solely originates from
$n_f = 5$ flavors of light quarks and gluons, and any effect induced by
the massive top and stop fields is decoupled and absorbed in the
renormalization constant of $\alpha_s$,
\be\bsp
  \frac{\delta\alpha_s}{\alpha_s} =&\
    \frac{\alpha_s}{2\pi\bar\epsilon} \bigg[\frac{n_f}{3} - \frac{11}{2}\bigg] +
    \frac{\alpha_s}{6\pi}
      \bigg[\frac{1}{\bar\epsilon} - \log\frac{m_t^2}{\mu_R^2}\bigg] \\ & \ +
    \frac{\alpha_s}{24\pi}
      \bigg[\frac{1}{\bar\epsilon} - \log\frac{m_3^2}{\mu_R^2}\bigg] \ .
\esp\ee

All loop-calculations achieved in this work rely on the OPP formalism. It
is based on the decomposition of any loop amplitude in both cut-constructible
and rational elements, the latter being
related to the $\epsilon$-pieces of the loop-integral
denominators ($R_1$) and numerators ($R_2$). For any renormalizable theory, there is
a finite number of $R_2$ terms, and they all involve interactions with at most four external
legs that can be seen as counterterms derived from the
tree-level Lagrangian~\cite{Ossola:2008xq}.
Considering corrections at the first order in QCD, the
$\sigma_3$-field induces three additional $R_2$ counterterms with
respect to the Standard Model case,
\be\bsp
 R_2^{\sigma_3^\dag\sigma_3} = &\ \frac{i g_s^2}{72 \pi^2} \delta_{c_1c_2}\Big[ 3 m_3^2 - p^2 \Big]\ , \\
 R_2^{g\sigma_3^\dag\sigma_3} = &\ \frac{53 i g_s^3}{576 \pi^2} T^{a_1}_{c_2c_3} \big(p_2-p_3\big)^{\mu_1} \ , \\
 R_2^{gg\sigma_3^\dag\sigma_3} = &\ \frac{i g_s^4}{1152\pi^2} \eta^{\mu_1\mu_2}
   \big[ 3\delta^{a_1a_2} - 187 \{T^{a_1},T^{a_2}\}\big]_{c_3c_4}\ , \\
\esp\ee
where $c_i$, $\mu_i$, and $p_i$ indicate the color index,
Lorentz index, and the four-momentum of the $i^{\rm th}$ particle incoming to the
$R_2^{\ldots i\ldots}$ vertex, respectively.
Moreover, the matrices $T$ denote fundamental representation matrices of $SU(3)$.

Contrary to complete supersymmetric scenarios, the $\tilde g$ operators
present a non-trivial one-loop ultraviolet behavior
that is not compensated by effects of other fields
such as gluinos. Since we focus on QCD NLO corrections to the strong production
of a pair of $\sigma_3$ fields followed by their LO decays, the
related counterterms are therefore omitted from this document.

Our stop simplified model has been implemented in \fr, and we have
employed the \nloct\ package to automatically generate all QCD ultraviolet and $R_2$
counterterms (including the Standard Model ones). The output
has been validated against our analytical calculations, which constitutes
a validation of the handling of new massive colored states by
\nloct. Finally,
the analytical results have been
exported to a UFO module that we have imported into \amc.
For our numerical analysis, we consider scenarios where
$m_3$ and $m_\chi$ are kept free. The $\tilde g_{L,R}$ parameters are
fixed to typical values for
supersymmetric models featuring a bino-like neutralino and a maximally-mixing
top squark,
\be
  \tilde g_L = 0.25 \qquad\text{and}\qquad
  \tilde g_R = 0.06 \ .
\ee

\begin{table*}[t]
 \begin{center}
   \begin{tabular}{c||c|c||c|c}
       \multirow{2}{*}{$m_3$ [GeV]} &
       \multicolumn{2}{|c||}{8 TeV} &
       \multicolumn{2}{|c}{13 TeV}\\
          & $\sigma^{\rm LO}$ [pb] & $\sigma^{\rm NLO}$ [pb] &
          $\sigma^{\rm LO}$ [pb] & $\sigma^{\rm NLO}$ [pb] \\
     \hline&&&&\\[-.4cm] 100 &
          $389.3^{\sss +34.2\%}_{\sss -23.9\%}$ &
          $554.8^{\sss +14.9\%}_{\sss -13.5\%}{}^{\sss +1.6\%}_{\sss -1.6\%}$ &
          $1066^{\sss +29.1\%}_{\sss -21.4\%}$ &
          $1497^{\sss +14.1\%}_{\sss -12.1\%}{}^{\sss +1.2\%}_{\sss -1.2\%}$\\[.05cm]
      250 &
          $4.118^{\sss +40.4\%}_{\sss -27.2\%}$ &
          $5.503^{\sss +13.1\%}_{\sss -13.7\%}{}^{\sss +3.7\%}_{\sss -3.7\%}$&
          $15.53^{\sss +35.2\%}_{\sss -24.8\%}$ &
          $21.56^{\sss +12.1\%}_{\sss -12.3\%}{}^{\sss +2.4\%}_{\sss -2.4\%}$ \\[.05cm]
      500 &
          $\big(6.594\times10^{-2}\big){}^{\sss +45.5\%}_{\sss -29.1\%}$ &
          $\big(7.764\times10^{-2}\big){}^{\sss +12.1\%}_{\sss -14.1\%}{}^{\sss+6.7\%}_{\sss -6.7\%}$ &
          $0.3890^{\sss +39.6\%}_{\sss -26.4\%}$ &
          $0.5062^{\sss +11.2\%}_{\sss -12.8\%}{}^{\sss +4.4\%}_{\sss -4.4\%}$ \\[.05cm]
      750 &
          $\big(3.504\times10^{-3}\big){}^{\sss +48.8\%}_{\sss -30.5\%}$ &
          $\big(3.699\times10^{-3}\big){}^{\sss +12.3\%}_{\sss -14.6\%}{}^{\sss+10.2\%}_{\sss-10.2\%}$&
          $\big(3.306\times10^{-2}\big){}^{\sss +41.8\%}_{\sss -27.5\%}$ &
          $\big(4.001\times10^{-2}\big){}^{\sss +10.8\%}_{\sss -12.9\%}{}^{\sss +6.1\%}_{\sss -6.1\%}$ \\[.05cm]
      1000&
          $\big(2.875\times10^{-4}\big){}^{\sss +51.5\%}_{\sss -31.5\%}$ &
          $\big(2.775\times10^{-4}\big){}^{\sss +13.1\%}_{\sss -15.2\%}{}^{\sss +15.5\%}_{\sss -15.5\%}$&
          $\big(4.614\times10^{-3}\big){}^{\sss +43.6\%}_{\sss -28.3\%}$ &
          $\big(5.219\times10^{-3}\big){}^{\sss +10.9\%}_{\sss -13.2\%}{}^{\sss +7.9\%}_{\sss -7.9\%}$\\
    \end{tabular}\\[.04cm]

    \begin{tabular}{c||c|c||c|c}
       \multirow{2}{*}{$m_8$ [GeV]} &
       \multicolumn{2}{|c||}{8 TeV} &
       \multicolumn{2}{|c}{13 TeV}\\
          & $\sigma^{\rm LO}$ [pb] & $\sigma^{\rm NLO}$ [pb] &
          $\sigma^{\rm LO}$ [pb] & $\sigma^{\rm NLO}$ [pb] \\
     \hline&&&&\\[-.4cm] 100 &
          $3854^{\sss +34.4\%}_{\sss -24.1\%}$ &
          $5573^{\sss +14.9\%}_{\sss -13.6\%}{}^{\sss +1.6\%}_{\sss -1.6\%}$&
          $10560^{\sss +29.2\%}_{\sss -21.5\%}$ &
          $14700^{\sss +13.6\%}_{\sss -11.9\%}{}^{\sss +1.2\%}_{\sss-1.2\%}$\\[.05cm]
      250 &
          $38.89^{\sss +41.3\%}_{\sss -27.7\%}$&
          $54.32^{\sss +14.5\%}_{\sss -14.6\%}{}^{\sss +3.9\%}_{\sss -3.9\%}$ &
          $150.4^{\sss +35.7\%}_{\sss -25.1\%}$&
          $214.5^{\sss +12.9\%}_{\sss -12.9\%}{}^{\sss +2.5\%}_{\sss -2.5\%}$ \\[.05cm]
      500 &
          $0.5878^{\sss +47.6\%}_{\sss -30.0\%}$&
          $0.7431^{\sss +15.8\%}_{\sss -16.2\%}{}^{\sss +7.6\%}_{\sss -7.6\%}$&
          $3.619^{\sss +40.8\%}_{\sss -27.0\%}$ &
          $4.977^{\sss +13.3\%}_{\sss -14.1\%}{}^{\sss +4.7\%}_{\sss -4.7\%}$ \\[.05cm]
      750 &
          $\big(2.977\times10^{-2}\big){}^{\sss +52.0\%}_{\sss -31.9\%}$ &
          $\big(3.353\times10^{-2}\big){}^{\sss +17.2\%}_{\sss -17.3\%}{}^{\sss +12.1\%}_{\sss -12.1\%}$&
          $0.2951^{\sss +43.6\%}_{\sss -28.4\%}$ &
          $0.3817^{\sss +14.0\%}_{\sss -14.8\%}{}^{\sss +6.9\%}_{\sss -6.9\%}$ \\[.05cm]
      1000&
          $\big(2.328\times10^{-3}\big){}^{\sss +55.9\%}_{\sss -33.4\%}$ &
          $\big(2.398\times10^{-3}\big){}^{\sss +19.0\%}_{\sss -18.4\%}{}^{\sss +19.1\%}_{\sss -19.1\%}$&
          $\big(3.983\times10^{-2}\big){}^{\sss +46.1\%}_{\sss -29.5\%}$ &
          $\big(4.822\times10^{-2}\big){}^{\sss +15.1\%}_{\sss -15.6\%}{}^{\sss +9.3\%}_{\sss -9.3\%}$\\
\end{tabular}
\end{center}
 \caption{\small \label{tab:xsec}Total cross sections for stop (upper panel)
   and sgluon (lower panel) pair production
   at the LHC, running at $\sqrt{s} = 8$ and 13~TeV. Results are
   presented together with the associated scale and PDF (not shown for
   the LO case) uncertainties. Monte Carlo errors are of about 0.2-0.3\% and omitted.}
\end{table*}

\section{Benchmark scenarios for sgluon hadroproduction}
We construct a simplified model describing sgluon dynamics by
supplementing the Standard Model with a real scalar field $\sigma_8$ (a sgluon)
of mass $m_8$ lying in the adjoint representation of the QCD gauge group.
Its strong interactions are described by gauge-covariant kinetic
terms and we enable single sgluon couplings to quarks and
gluons, like in complete models where such interactions are
loop-induced. The corresponding effective Lagrangian reads
\be\bsp
  {\cal L}_8 =&\
    \frac12 D_\mu \sigma_8 D^\mu \sigma_8 - \frac12 m_8^2 \sigma_8 \sigma_8
   + \frac{{\hat g}_g}{\Lambda} \sigma_8 G_{\mu\nu} G^{\mu\nu}  \\
   &\quad + \sum_{q=u,d} \Big[\sigma_8 \bar q \big(\hat g^L_q P_L+\hat g^R_q P_R\big)
       q + {\rm h.c.} \Big]\ ,
\esp\ee
where $G^{\mu\nu}$ refers to the gluon
field strength tensor and the single sgluon interaction strengths are denoted by
$\hat g$. Although the $\hat g$ operators induce single sgluon production,
we ignore it in this work since the presence of a \textit{complete} basis
of dimension-five operators at tree-level is required 
to guarantee the cancellation, after renormalization, of all
loop-induced ultraviolet divergences.
We postpone the associated study to a future work.

The ${\hat g}$ couplings being technically of higher-order in QCD (as in
complete theories), the quark fields are renormalized like in the
Standard Model. In contrast, the sgluon QCD interactions induce a
modification of the on-shell gluon wave-function renormalization constant
$\delta Z_g$ and
yield non-vanishing on-shell sgluon wave-function ($\delta Z_{\sigma_8}$)
and mass ($\delta m^2_8$) renormalization constants,
\be\bsp
  & \delta Z_g = \delta Z_g^{(SM)} - \frac{g_s^2}{32 \pi^2}
    \bigg[\frac{1}{\bar\epsilon} - \log\frac{m_8^2}{\mu_R^2}\bigg] \ , \\
  & \delta Z_{\sigma_8} = 0
  \quad\text{and}\quad
  \delta m_8^2 = -\frac{3 g_s^2 m_8^2}{16\pi^2}
      \Big[ \frac{3}{\bar\epsilon} + 7 - 3\log\frac{m_8^2}{\mu_R^2} \Big] \ .
\esp\ee
Sgluon effects are also subtracted, at zero-momentum transfer, from the
gluon self-energy and absorbed in the renormalization of the strong coupling,
\be\bsp
  \frac{\delta\alpha_s}{\alpha_s} =&\
    \frac{\alpha_s}{2\pi\bar\epsilon} \bigg[\frac{n_f}{3} - \frac{11}{2}\bigg] +
    \frac{\alpha_s}{6\pi}
      \bigg[\frac{1}{\bar\epsilon} - \log\frac{m_t^2}{\mu_R^2}\bigg] \\ & \ +
    \frac{\alpha_s}{8\pi}
      \bigg[\frac{1}{\bar\epsilon} - \log\frac{m_8^2}{\mu_R^2}\bigg] \ .
\esp\ee
They finally induce new $R_2$ counterterms,
\be\bsp
 R_2^{\sigma_8\sigma_8} = &\ \frac{i g_s^2}{32 \pi^2} \delta_{a_1a_2} \Big[ 3 m_8^2 - p^2 \Big]\ , \\
 R_2^{g\sigma_8\sigma_8} = &\ \frac{7 g_s^3}{64 \pi^2} f_{a_1 a_2 a_3} \big(p_2-p_3\big)^{\mu_1}\ , \\
 R_2^{gg\sigma_8\sigma_8} = &\ \frac{i g_s^4}{384 \pi^2} \eta^{\mu_1\mu_2}
   \Big[72 (d_{a_1a_4e} d_{a_2a_3e} +d_{a_1a_3e} d_{a_2a_4e})\\
     &\quad - 141  d_{a_1a_2e} d_{a_3a_4e} -92 \delta_{a_1a_2}\delta_{a_3a_4}\\
     &\quad + 50 (\delta_{a_1a_3}\delta_{a_2a_4} + \delta_{a_1a_4}\delta_{a_2a_3})\Big]\ ,
\esp\ee
in the same notations as in the previous section.

We have implemented the sgluon simplified model in
\fr\ and generated a UFO model that we have linked to \amc\ by means of the
\nloct\ package. The generated model has then been validated analytically against the
above results.
Our numerical study relies on benchmark scenarios inspired by an
$R$-symmetric supersymmetric setup with
non-minimal flavor violation in the squark sector~\cite{Calvet:2012rk}, in which
the only non-vanishing coupling parameters are fixed to
\be\bsp
  & \frac{\hat g_g}{\Lambda} = 1.5\cdot 10^{-6}~\text{GeV}^{-1}\ ,\\
  & (\hat g_u^{L,R})_{3i} = (\hat g_u^{L,R})_{i3} = 3\cdot 10^{-3}\quad
     \forall i=1,2,3\ .
\esp\ee

\section{LHC phenomenology}
In Tab.~\ref{tab:xsec}, we provide stop and sgluon pair production cross
sections for LHC collisions at center-of-mass energies of
$\sqrt{s} = $ 8 and
13~TeV and for different mass choices. The results are
evaluated both at the LO and NLO accuracy and presented together with the
associated theoretical uncertainties. For the central values,
we have fixed the renormalization and factorization scales to the stop/sgluon
mass and used the NNPDF~2.3 parton
distributions~\cite{Ball:2012cx}. Scale uncertainties have been derived
by varying both scales by a factor of two up
and down, and the parton distribution uncertainties have been extracted from the
cross section values spanned by the NNPDF density replica.

The results of Tab.~\ref{tab:xsec} have been confronted to predictions
obtained with the public packages
{\sc Prospino}~\cite{Beenakker:1997ut} (stop pair production) and
{\sc MadGolem}~\cite{GoncalvesNetto:2012nt} (sgluon pair production).
Stop-pair total production rates
have been found to agree at the level of the numerical integration
error, while virtual and real contributions to sgluon-pair
production are agreeing separately at the amplitude level.
We have additionally
performed independent calculations of the loop contributions based on
{\sc FeynArts}~\cite{Hahn:2000kx},
that we have found to agree with our predictions.

\begin{figure}
\centering
  \includegraphics[width=.90\columnwidth]{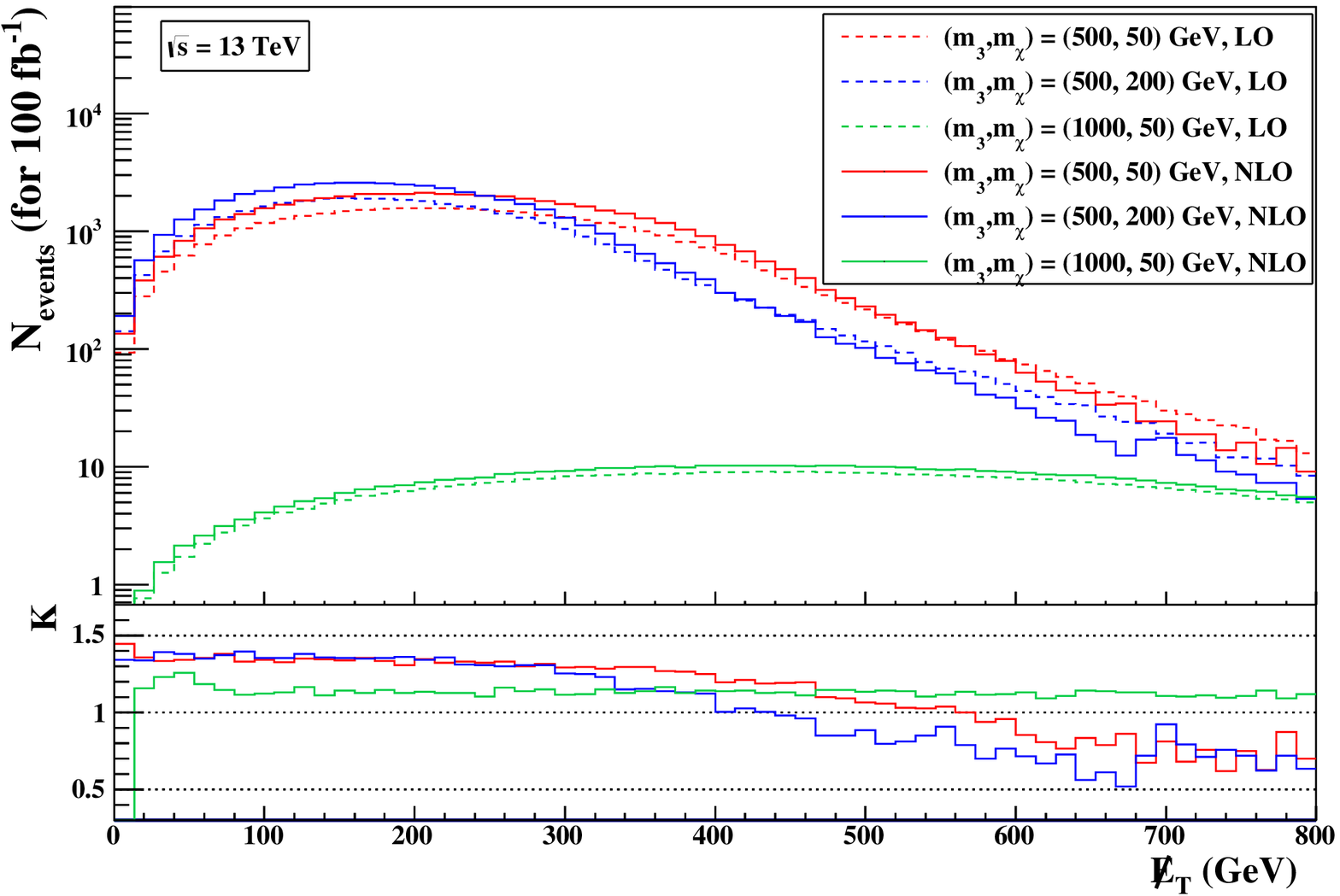}
  \caption{\small \label{fig:met_s3}Missing transverse energy spectrum for
    a stop pair production and decay signal. We consider several mass setups
    and show results at the NLO and LO accuracy (upper panel),
    together with their bin-by-bin ratio (lower panel).}
  \includegraphics[width=.90\columnwidth]{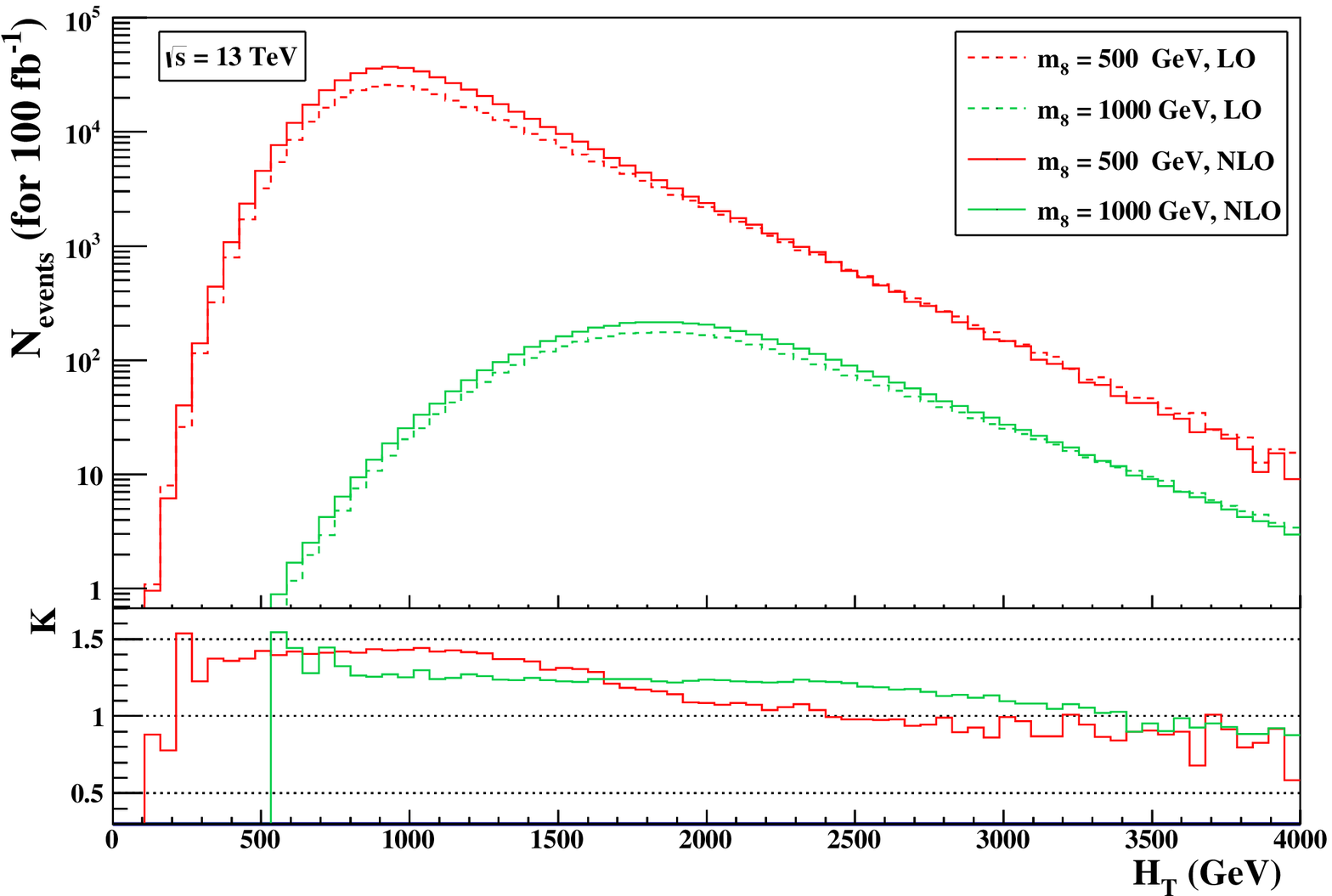}
  \caption{\small \label{fig:ht_s8} Same as Fig.~\ref{fig:met_s3} for the
    $H_T$ spectrum of sgluon signals.}
\end{figure}

Realistic descriptions of LHC collisions require
to match hard scattering matrix elements
to a modeling of QCD environment. To this aim, we make
use of the {\sc MC@NLO} method~\cite{Frixione:2002ik} as implemented
in \amc. We match in this way the hard scattering process to
the {\sc Pythia}~8 parton
showering and hadronization~\cite{Sjostrand:2007gs}, after employing the {\sc MadSpin} and
{\sc MadWidth} programs to handle stop and sgluon decays.
Jet reconstruction is then performed by means of the
anti-$k_T$ algorithm with a radius parameter set to
0.4~\cite{Cacciari:2008gp}, as included in the {\sc FastJet}
program~\cite{Cacciari:2011ma}, and events are finally analyzed
with the {\sc MadAnalysis}~5 package~\cite{Conte:2012fm}.
Normalizing the results to an integrated luminosity of 100~fb$^{-1}$,
we present, in Fig.~\ref{fig:met_s3}, the distribution of a key observable
for stop searches, namely the missing transverse energy
We show LO and NLO predictions for 13~TeV collisions as calculated by \amc\
in the context of three benchmark scenarios for which
$(m_3,m_\chi) = (500,50)$~GeV (red), $(1000,50)$~GeV
(green) and $(500,200)$~GeV (blue). Similarly, we describe the hadronic
activity $H_T$ associated with the production of a sgluon pair in
Fig.~\ref{fig:ht_s8} in the case of
a sgluon mass of 500~GeV (red) and 1000~GeV (green)\footnote{
With the current level of precision of the experimental searches, the current
limits on the stop and sgluon masses, that are
extracted on the basis of NLO total rates but LO simulations for the distributions,
can be assumed to hold. Evaluating the NLO effects on the shapes and how this
translates in terms of a modification of the limits goes beyond the scope of this
work.}.

\section{Conclusions}
In this Letter, we have demonstrated that a joint use of the \fr, \nloct\ and \amc\
programs enables the full automation of the Monte Carlo simulations of high-energy
physics collisions at the
next-to-leading order accuracy in QCD and for non-trivial extensions of the Standard Model. This has
been illustrated with simplified models such as those used for supersymmetry searches
at the LHC. In this context, we have adopted setups that exhibit extra colored particles and non-usual
interaction structures and presented the analysis of two exemplary signals
with the automated tool \ma.

In the aim of an embedding within experimental software, we have
designed a webpage,
\verb+ http://feynrules.irmp.ucl.ac.be/wiki/NLOModels+
where hundreds of differential distributions are available
for validation purposes, together with the associated
\fr\ and UFO models.

\acknowledgments
We are extremely grateful to D.~Goncalves-Netto,
D.~Lopez-Val and K.~Mawatari
for their help with {\sc MadGolem}. We also thank R.~Frederix,
S.~Frixione,
F.~Maltoni, O.~Mattelaer, P.~Torrielli and M.~Zaro for enlightening
discussions. This work has been supported in part by the
ERC grant 291377 \textit{LHCtheory: Theoretical predictions and analyses of LHC physics:
advancing the precision frontier}, the Research Executive
Agency of the European Union under Grant Agreement
PITN-GA-2012-315877 (MCNet) and the Theory-LHC-France initiative of the CNRS/IN2P3.
CD is a Durham International Junior Research Fellow, the work of VH is supported by
the SNF with grant PBELP2 146525 and the one of
JP by a PhD grant of the \textit{Investissements d'avenir,
Labex ENIGMASS}.

\bibliography{biblio}

\begin{thebibliography}{26}%
\makeatletter
\providecommand \@ifxundefined [1]{%
 \@ifx{#1\undefined}
}%
\providecommand \@ifnum [1]{%
 \ifnum #1\expandafter \@firstoftwo
 \else \expandafter \@secondoftwo
 \fi
}%
\providecommand \@ifx [1]{%
 \ifx #1\expandafter \@firstoftwo
 \else \expandafter \@secondoftwo
 \fi
}%
\providecommand \natexlab [1]{#1}%
\providecommand \enquote  [1]{``#1''}%
\providecommand \bibnamefont  [1]{#1}%
\providecommand \bibfnamefont [1]{#1}%
\providecommand \citenamefont [1]{#1}%
\providecommand \href@noop [0]{\@secondoftwo}%
\providecommand \href [0]{\begingroup \@sanitize@url \@href}%
\providecommand \@href[1]{\@@startlink{#1}\@@href}%
\providecommand \@@href[1]{\endgroup#1\@@endlink}%
\providecommand \@sanitize@url [0]{\catcode `\\12\catcode `\$12\catcode
  `\&12\catcode `\#12\catcode `\^12\catcode `\_12\catcode `\%12\relax}%
\providecommand \@@startlink[1]{}%
\providecommand \@@endlink[0]{}%
\providecommand \url  [0]{\begingroup\@sanitize@url \@url }%
\providecommand \@url [1]{\endgroup\@href {#1}{\urlprefix }}%
\providecommand \urlprefix  [0]{URL }%
\providecommand \Eprint [0]{\href }%
\providecommand \doibase [0]{http://dx.doi.org/}%
\providecommand \selectlanguage [0]{\@gobble}%
\providecommand \bibinfo  [0]{\@secondoftwo}%
\providecommand \bibfield  [0]{\@secondoftwo}%
\providecommand \translation [1]{[#1]}%
\providecommand \BibitemOpen [0]{}%
\providecommand \bibitemStop [0]{}%
\providecommand \bibitemNoStop [0]{.\EOS\space}%
\providecommand \EOS [0]{\spacefactor3000\relax}%
\providecommand \BibitemShut  [1]{\csname bibitem#1\endcsname}%
\let\auto@bib@innerbib\@empty
\bibitem [{\citenamefont {Nilles}(1984)}]{Nilles:1983ge}%
  \BibitemOpen
  \bibfield  {author} {\bibinfo {author} {\bibfnamefont {H.~P.}\ \bibnamefont
  {Nilles}},\ }\href {\doibase 10.1016/0370-1573(84)90008-5} {\bibfield
  {journal} {\bibinfo  {journal} {Phys.Rept.}\ }\textbf {\bibinfo {volume}
  {110}},\ \bibinfo {pages} {1} (\bibinfo {year} {1984})}\BibitemShut {NoStop}%
\bibitem [{\citenamefont {Haber}\ and\ \citenamefont
  {Kane}(1985)}]{Haber:1984rc}%
  \BibitemOpen
  \bibfield  {author} {\bibinfo {author} {\bibfnamefont {H.~E.}\ \bibnamefont
  {Haber}}\ and\ \bibinfo {author} {\bibfnamefont {G.~L.}\ \bibnamefont
  {Kane}},\ }\href {\doibase 10.1016/0370-1573(85)90051-1} {\bibfield
  {journal} {\bibinfo  {journal} {Phys.Rept.}\ }\textbf {\bibinfo {volume}
  {117}},\ \bibinfo {pages} {75} (\bibinfo {year} {1985})}\BibitemShut
  {NoStop}%
\bibitem [{\citenamefont {Salam}\ and\ \citenamefont
  {Strathdee}(1975)}]{Salam:1974xa}%
  \BibitemOpen
  \bibfield  {author} {\bibinfo {author} {\bibfnamefont {A.}~\bibnamefont
  {Salam}}\ and\ \bibinfo {author} {\bibfnamefont {J.}~\bibnamefont
  {Strathdee}},\ }\href {\doibase 10.1016/0550-3213(75)90253-9} {\bibfield
  {journal} {\bibinfo  {journal} {Nucl.Phys.}\ }\textbf {\bibinfo {volume}
  {B87}},\ \bibinfo {pages} {85} (\bibinfo {year} {1975})}\BibitemShut
  {NoStop}%
\bibitem [{\citenamefont {Fayet}(1976)}]{Fayet:1975yi}%
  \BibitemOpen
  \bibfield  {author} {\bibinfo {author} {\bibfnamefont {P.}~\bibnamefont
  {Fayet}},\ }\href {\doibase 10.1016/0550-3213(76)90458-2} {\bibfield
  {journal} {\bibinfo  {journal} {Nucl.Phys.}\ }\textbf {\bibinfo {volume}
  {B113}},\ \bibinfo {pages} {135} (\bibinfo {year} {1976})}\BibitemShut
  {NoStop}%
\bibitem [{\citenamefont {Kilic}\ \emph {et~al.}(2010)\citenamefont {Kilic},
  \citenamefont {Okui},\ and\ \citenamefont {Sundrum}}]{Kilic:2009mi}%
  \BibitemOpen
  \bibfield  {author} {\bibinfo {author} {\bibfnamefont {C.}~\bibnamefont
  {Kilic}}, \bibinfo {author} {\bibfnamefont {T.}~\bibnamefont {Okui}}, \ and\
  \bibinfo {author} {\bibfnamefont {R.}~\bibnamefont {Sundrum}},\ }\href
  {\doibase 10.1007/JHEP02(2010)018} {\bibfield  {journal} {\bibinfo  {journal}
  {JHEP}\ }\textbf {\bibinfo {volume} {1002}},\ \bibinfo {pages} {018}
  (\bibinfo {year} {2010})}\BibitemShut {NoStop}%
\bibitem [{\citenamefont {Alwall}\ \emph
  {et~al.}(2014{\natexlab{a}})\citenamefont {Alwall}, \citenamefont {Frederix},
  \citenamefont {Frixione}, \citenamefont {Hirschi}, \citenamefont {Maltoni}
  \emph {et~al.}}]{Alwall:2014hca}%
  \BibitemOpen
  \bibfield  {author} {\bibinfo {author} {\bibfnamefont {J.}~\bibnamefont
  {Alwall}}, \bibinfo {author} {\bibfnamefont {R.}~\bibnamefont {Frederix}},
  \bibinfo {author} {\bibfnamefont {S.}~\bibnamefont {Frixione}}, \bibinfo
  {author} {\bibfnamefont {V.}~\bibnamefont {Hirschi}}, \bibinfo {author}
  {\bibfnamefont {F.}~\bibnamefont {Maltoni}},  \emph {et~al.},\ }\href
  {\doibase 10.1007/JHEP07(2014)079} {\bibfield  {journal} {\bibinfo  {journal}
  {JHEP}\ }\textbf {\bibinfo {volume} {1407}},\ \bibinfo {pages} {079}
  (\bibinfo {year} {2014}{\natexlab{a}})}\BibitemShut {NoStop}%
\bibitem [{\citenamefont {Degrande}(2014)}]{Degrande:2014vpa}%
  \BibitemOpen
  \bibfield  {author} {\bibinfo {author} {\bibfnamefont {C.}~\bibnamefont
  {Degrande}},\ }\href@noop {} {\  (\bibinfo {year} {2014})},\ \Eprint
  {http://arxiv.org/abs/1406.3030} {arXiv:1406.3030 [hep-ph]} \BibitemShut
  {NoStop}%
\bibitem [{\citenamefont {Alloul}\ \emph {et~al.}(2014)\citenamefont {Alloul},
  \citenamefont {Christensen}, \citenamefont {Degrande}, \citenamefont {Duhr},\
  and\ \citenamefont {Fuks}}]{Alloul:2013bka}%
  \BibitemOpen
  \bibfield  {author} {\bibinfo {author} {\bibfnamefont {A.}~\bibnamefont
  {Alloul}}, \bibinfo {author} {\bibfnamefont {N.~D.}\ \bibnamefont
  {Christensen}}, \bibinfo {author} {\bibfnamefont {C.}~\bibnamefont
  {Degrande}}, \bibinfo {author} {\bibfnamefont {C.}~\bibnamefont {Duhr}}, \
  and\ \bibinfo {author} {\bibfnamefont {B.}~\bibnamefont {Fuks}},\ }\href
  {\doibase 10.1016/j.cpc.2014.04.012} {\bibfield  {journal} {\bibinfo
  {journal} {Comput.Phys.Commun.}\ }\textbf {\bibinfo {volume} {185}},\
  \bibinfo {pages} {2250} (\bibinfo {year} {2014})}\BibitemShut {NoStop}%
\bibitem [{\citenamefont {Alwall}\ \emph {et~al.}(2009)\citenamefont {Alwall},
  \citenamefont {Schuster},\ and\ \citenamefont {Toro}}]{Alwall:2008ag}%
  \BibitemOpen
  \bibfield  {author} {\bibinfo {author} {\bibfnamefont {J.}~\bibnamefont
  {Alwall}}, \bibinfo {author} {\bibfnamefont {P.}~\bibnamefont {Schuster}}, \
  and\ \bibinfo {author} {\bibfnamefont {N.}~\bibnamefont {Toro}},\ }\href
  {\doibase 10.1103/PhysRevD.79.075020} {\bibfield  {journal} {\bibinfo
  {journal} {Phys.Rev.}\ }\textbf {\bibinfo {volume} {D79}},\ \bibinfo {pages}
  {075020} (\bibinfo {year} {2009})}\BibitemShut {NoStop}%
\bibitem [{\citenamefont {Alves}\ \emph {et~al.}(2012)\citenamefont {Alves}
  \emph {et~al.}}]{Alves:2011wf}%
  \BibitemOpen
  \bibfield  {author} {\bibinfo {author} {\bibfnamefont {D.}~\bibnamefont
  {Alves}} \emph {et~al.} (\bibinfo {collaboration} {LHC New Physics Working
  Group}),\ }\href {\doibase 10.1088/0954-3899/39/10/105005} {\bibfield
  {journal} {\bibinfo  {journal} {J.Phys.}\ }\textbf {\bibinfo {volume}
  {G39}},\ \bibinfo {pages} {105005} (\bibinfo {year} {2012})}\BibitemShut
  {NoStop}%
\bibitem [{\citenamefont {Degrande}\ \emph {et~al.}(2012)\citenamefont
  {Degrande}, \citenamefont {Duhr}, \citenamefont {Fuks}, \citenamefont
  {Grellscheid}, \citenamefont {Mattelaer} \emph {et~al.}}]{Degrande:2011ua}%
  \BibitemOpen
  \bibfield  {author} {\bibinfo {author} {\bibfnamefont {C.}~\bibnamefont
  {Degrande}}, \bibinfo {author} {\bibfnamefont {C.}~\bibnamefont {Duhr}},
  \bibinfo {author} {\bibfnamefont {B.}~\bibnamefont {Fuks}}, \bibinfo {author}
  {\bibfnamefont {D.}~\bibnamefont {Grellscheid}}, \bibinfo {author}
  {\bibfnamefont {O.}~\bibnamefont {Mattelaer}},  \emph {et~al.},\ }\href
  {\doibase 10.1016/j.cpc.2012.01.022} {\bibfield  {journal} {\bibinfo
  {journal} {Comput.Phys.Commun.}\ }\textbf {\bibinfo {volume} {183}},\
  \bibinfo {pages} {1201} (\bibinfo {year} {2012})}\BibitemShut {NoStop}%
\bibitem [{\citenamefont {Hirschi}\ \emph {et~al.}(2011)\citenamefont
  {Hirschi}, \citenamefont {Frederix}, \citenamefont {Frixione}, \citenamefont
  {Garzelli}, \citenamefont {Maltoni} \emph {et~al.}}]{Hirschi:2011pa}%
  \BibitemOpen
  \bibfield  {author} {\bibinfo {author} {\bibfnamefont {V.}~\bibnamefont
  {Hirschi}}, \bibinfo {author} {\bibfnamefont {R.}~\bibnamefont {Frederix}},
  \bibinfo {author} {\bibfnamefont {S.}~\bibnamefont {Frixione}}, \bibinfo
  {author} {\bibfnamefont {M.~V.}\ \bibnamefont {Garzelli}}, \bibinfo {author}
  {\bibfnamefont {F.}~\bibnamefont {Maltoni}},  \emph {et~al.},\ }\href
  {\doibase 10.1007/JHEP05(2011)044} {\bibfield  {journal} {\bibinfo  {journal}
  {JHEP}\ }\textbf {\bibinfo {volume} {1105}},\ \bibinfo {pages} {044}
  (\bibinfo {year} {2011})}\BibitemShut {NoStop}%
\bibitem [{\citenamefont {Ossola}\ \emph {et~al.}(2007)\citenamefont {Ossola},
  \citenamefont {Papadopoulos},\ and\ \citenamefont {Pittau}}]{Ossola:2006us}%
  \BibitemOpen
  \bibfield  {author} {\bibinfo {author} {\bibfnamefont {G.}~\bibnamefont
  {Ossola}}, \bibinfo {author} {\bibfnamefont {C.~G.}\ \bibnamefont
  {Papadopoulos}}, \ and\ \bibinfo {author} {\bibfnamefont {R.}~\bibnamefont
  {Pittau}},\ }\href {\doibase 10.1016/j.nuclphysb.2006.11.012} {\bibfield
  {journal} {\bibinfo  {journal} {Nucl.Phys.}\ }\textbf {\bibinfo {volume}
  {B763}},\ \bibinfo {pages} {147} (\bibinfo {year} {2007})}\BibitemShut
  {NoStop}%
\bibitem [{\citenamefont {Artoisenet}\ \emph {et~al.}(2013)\citenamefont
  {Artoisenet}, \citenamefont {Frederix}, \citenamefont {Mattelaer},\ and\
  \citenamefont {Rietkerk}}]{Artoisenet:2012st}%
  \BibitemOpen
  \bibfield  {author} {\bibinfo {author} {\bibfnamefont {P.}~\bibnamefont
  {Artoisenet}}, \bibinfo {author} {\bibfnamefont {R.}~\bibnamefont
  {Frederix}}, \bibinfo {author} {\bibfnamefont {O.}~\bibnamefont {Mattelaer}},
  \ and\ \bibinfo {author} {\bibfnamefont {R.}~\bibnamefont {Rietkerk}},\
  }\href {\doibase 10.1007/JHEP03(2013)015} {\bibfield  {journal} {\bibinfo
  {journal} {JHEP}\ }\textbf {\bibinfo {volume} {1303}},\ \bibinfo {pages}
  {015} (\bibinfo {year} {2013})}\BibitemShut {NoStop}%
\bibitem [{\citenamefont {Alwall}\ \emph
  {et~al.}(2014{\natexlab{b}})\citenamefont {Alwall}, \citenamefont {Duhr},
  \citenamefont {Fuks}, \citenamefont {Mattelaer}, \citenamefont {Ozturk} \emph
  {et~al.}}]{Alwall:2014bza}%
  \BibitemOpen
  \bibfield  {author} {\bibinfo {author} {\bibfnamefont {J.}~\bibnamefont
  {Alwall}}, \bibinfo {author} {\bibfnamefont {C.}~\bibnamefont {Duhr}},
  \bibinfo {author} {\bibfnamefont {B.}~\bibnamefont {Fuks}}, \bibinfo {author}
  {\bibfnamefont {O.}~\bibnamefont {Mattelaer}}, \bibinfo {author}
  {\bibfnamefont {D.~G.}\ \bibnamefont {Ozturk}},  \emph {et~al.},\ }\href@noop
  {} {\  (\bibinfo {year} {2014}{\natexlab{b}})},\ \Eprint
  {http://arxiv.org/abs/1402.1178} {arXiv:1402.1178 [hep-ph]} \BibitemShut
  {NoStop}%
\bibitem [{\citenamefont {Ossola}\ \emph {et~al.}(2008)\citenamefont {Ossola},
  \citenamefont {Papadopoulos},\ and\ \citenamefont {Pittau}}]{Ossola:2008xq}%
  \BibitemOpen
  \bibfield  {author} {\bibinfo {author} {\bibfnamefont {G.}~\bibnamefont
  {Ossola}}, \bibinfo {author} {\bibfnamefont {C.~G.}\ \bibnamefont
  {Papadopoulos}}, \ and\ \bibinfo {author} {\bibfnamefont {R.}~\bibnamefont
  {Pittau}},\ }\href {\doibase 10.1088/1126-6708/2008/05/004} {\bibfield
  {journal} {\bibinfo  {journal} {JHEP}\ }\textbf {\bibinfo {volume} {0805}},\
  \bibinfo {pages} {004} (\bibinfo {year} {2008})}\BibitemShut {NoStop}%
\bibitem [{\citenamefont {Calvet}\ \emph {et~al.}(2013)\citenamefont {Calvet},
  \citenamefont {Fuks}, \citenamefont {Gris},\ and\ \citenamefont
  {Valery}}]{Calvet:2012rk}%
  \BibitemOpen
  \bibfield  {author} {\bibinfo {author} {\bibfnamefont {S.}~\bibnamefont
  {Calvet}}, \bibinfo {author} {\bibfnamefont {B.}~\bibnamefont {Fuks}},
  \bibinfo {author} {\bibfnamefont {P.}~\bibnamefont {Gris}}, \ and\ \bibinfo
  {author} {\bibfnamefont {L.}~\bibnamefont {Valery}},\ }\href {\doibase
  10.1007/JHEP04(2013)043} {\bibfield  {journal} {\bibinfo  {journal} {JHEP}\
  }\textbf {\bibinfo {volume} {1304}},\ \bibinfo {pages} {043} (\bibinfo {year}
  {2013})}\BibitemShut {NoStop}%
\bibitem [{\citenamefont {Ball}\ \emph {et~al.}(2013)\citenamefont {Ball},
  \citenamefont {Bertone}, \citenamefont {Carrazza}, \citenamefont {Deans},
  \citenamefont {Del~Debbio} \emph {et~al.}}]{Ball:2012cx}%
  \BibitemOpen
  \bibfield  {author} {\bibinfo {author} {\bibfnamefont {R.~D.}\ \bibnamefont
  {Ball}}, \bibinfo {author} {\bibfnamefont {V.}~\bibnamefont {Bertone}},
  \bibinfo {author} {\bibfnamefont {S.}~\bibnamefont {Carrazza}}, \bibinfo
  {author} {\bibfnamefont {C.~S.}\ \bibnamefont {Deans}}, \bibinfo {author}
  {\bibfnamefont {L.}~\bibnamefont {Del~Debbio}},  \emph {et~al.},\ }\href
  {\doibase 10.1016/j.nuclphysb.2012.10.003} {\bibfield  {journal} {\bibinfo
  {journal} {Nucl.Phys.}\ }\textbf {\bibinfo {volume} {B867}},\ \bibinfo
  {pages} {244} (\bibinfo {year} {2013})}\BibitemShut {NoStop}%
\bibitem [{\citenamefont {Beenakker}\ \emph {et~al.}(1998)\citenamefont
  {Beenakker}, \citenamefont {Kramer}, \citenamefont {Plehn}, \citenamefont
  {Spira},\ and\ \citenamefont {Zerwas}}]{Beenakker:1997ut}%
  \BibitemOpen
  \bibfield  {author} {\bibinfo {author} {\bibfnamefont {W.}~\bibnamefont
  {Beenakker}}, \bibinfo {author} {\bibfnamefont {M.}~\bibnamefont {Kramer}},
  \bibinfo {author} {\bibfnamefont {T.}~\bibnamefont {Plehn}}, \bibinfo
  {author} {\bibfnamefont {M.}~\bibnamefont {Spira}}, \ and\ \bibinfo {author}
  {\bibfnamefont {P.}~\bibnamefont {Zerwas}},\ }\href {\doibase
  10.1016/S0550-3213(98)00014-5} {\bibfield  {journal} {\bibinfo  {journal}
  {Nucl.Phys.}\ }\textbf {\bibinfo {volume} {B515}},\ \bibinfo {pages} {3}
  (\bibinfo {year} {1998})}\BibitemShut {NoStop}%
\bibitem [{\citenamefont {Goncalves-Netto}\ \emph {et~al.}(2012)\citenamefont
  {Goncalves-Netto}, \citenamefont {Lopez-Val}, \citenamefont {Mawatari},
  \citenamefont {Plehn},\ and\ \citenamefont
  {Wigmore}}]{GoncalvesNetto:2012nt}%
  \BibitemOpen
  \bibfield  {author} {\bibinfo {author} {\bibfnamefont {D.}~\bibnamefont
  {Goncalves-Netto}}, \bibinfo {author} {\bibfnamefont {D.}~\bibnamefont
  {Lopez-Val}}, \bibinfo {author} {\bibfnamefont {K.}~\bibnamefont {Mawatari}},
  \bibinfo {author} {\bibfnamefont {T.}~\bibnamefont {Plehn}}, \ and\ \bibinfo
  {author} {\bibfnamefont {I.}~\bibnamefont {Wigmore}},\ }\href {\doibase
  10.1103/PhysRevD.85.114024} {\bibfield  {journal} {\bibinfo  {journal}
  {Phys.Rev.}\ }\textbf {\bibinfo {volume} {D85}},\ \bibinfo {pages} {114024}
  (\bibinfo {year} {2012})}\BibitemShut {NoStop}%
\bibitem [{\citenamefont {Hahn}(2001)}]{Hahn:2000kx}%
  \BibitemOpen
  \bibfield  {author} {\bibinfo {author} {\bibfnamefont {T.}~\bibnamefont
  {Hahn}},\ }\href {\doibase 10.1016/S0010-4655(01)00290-9} {\bibfield
  {journal} {\bibinfo  {journal} {Comput.Phys.Commun.}\ }\textbf {\bibinfo
  {volume} {140}},\ \bibinfo {pages} {418} (\bibinfo {year}
  {2001})}\BibitemShut {NoStop}%
\bibitem [{\citenamefont {Frixione}\ and\ \citenamefont
  {Webber}(2002)}]{Frixione:2002ik}%
  \BibitemOpen
  \bibfield  {author} {\bibinfo {author} {\bibfnamefont {S.}~\bibnamefont
  {Frixione}}\ and\ \bibinfo {author} {\bibfnamefont {B.~R.}\ \bibnamefont
  {Webber}},\ }\href {\doibase 10.1088/1126-6708/2002/06/029} {\bibfield
  {journal} {\bibinfo  {journal} {JHEP}\ }\textbf {\bibinfo {volume} {0206}},\
  \bibinfo {pages} {029} (\bibinfo {year} {2002})}\BibitemShut {NoStop}%
\bibitem [{\citenamefont {Sjostrand}\ \emph {et~al.}(2008)\citenamefont
  {Sjostrand}, \citenamefont {Mrenna},\ and\ \citenamefont
  {Skands}}]{Sjostrand:2007gs}%
  \BibitemOpen
  \bibfield  {author} {\bibinfo {author} {\bibfnamefont {T.}~\bibnamefont
  {Sjostrand}}, \bibinfo {author} {\bibfnamefont {S.}~\bibnamefont {Mrenna}}, \
  and\ \bibinfo {author} {\bibfnamefont {P.~Z.}\ \bibnamefont {Skands}},\
  }\href {\doibase 10.1016/j.cpc.2008.01.036} {\bibfield  {journal} {\bibinfo
  {journal} {Comput.Phys.Commun.}\ }\textbf {\bibinfo {volume} {178}},\
  \bibinfo {pages} {852} (\bibinfo {year} {2008})}\BibitemShut {NoStop}%
\bibitem [{\citenamefont {Cacciari}\ \emph {et~al.}(2008)\citenamefont
  {Cacciari}, \citenamefont {Salam},\ and\ \citenamefont
  {Soyez}}]{Cacciari:2008gp}%
  \BibitemOpen
  \bibfield  {author} {\bibinfo {author} {\bibfnamefont {M.}~\bibnamefont
  {Cacciari}}, \bibinfo {author} {\bibfnamefont {G.~P.}\ \bibnamefont {Salam}},
  \ and\ \bibinfo {author} {\bibfnamefont {G.}~\bibnamefont {Soyez}},\ }\href
  {\doibase 10.1088/1126-6708/2008/04/063} {\bibfield  {journal} {\bibinfo
  {journal} {JHEP}\ }\textbf {\bibinfo {volume} {0804}},\ \bibinfo {pages}
  {063} (\bibinfo {year} {2008})}\BibitemShut {NoStop}%
\bibitem [{\citenamefont {Cacciari}\ \emph {et~al.}(2012)\citenamefont
  {Cacciari}, \citenamefont {Salam},\ and\ \citenamefont
  {Soyez}}]{Cacciari:2011ma}%
  \BibitemOpen
  \bibfield  {author} {\bibinfo {author} {\bibfnamefont {M.}~\bibnamefont
  {Cacciari}}, \bibinfo {author} {\bibfnamefont {G.~P.}\ \bibnamefont {Salam}},
  \ and\ \bibinfo {author} {\bibfnamefont {G.}~\bibnamefont {Soyez}},\ }\href
  {\doibase 10.1140/epjc/s10052-012-1896-2} {\bibfield  {journal} {\bibinfo
  {journal} {Eur.Phys.J.}\ }\textbf {\bibinfo {volume} {C72}},\ \bibinfo
  {pages} {1896} (\bibinfo {year} {2012})}\BibitemShut {NoStop}%
\bibitem [{\citenamefont {Conte}\ \emph {et~al.}(2013)\citenamefont {Conte},
  \citenamefont {Fuks},\ and\ \citenamefont {Serret}}]{Conte:2012fm}%
  \BibitemOpen
  \bibfield  {author} {\bibinfo {author} {\bibfnamefont {E.}~\bibnamefont
  {Conte}}, \bibinfo {author} {\bibfnamefont {B.}~\bibnamefont {Fuks}}, \ and\
  \bibinfo {author} {\bibfnamefont {G.}~\bibnamefont {Serret}},\ }\href
  {\doibase 10.1016/j.cpc.2012.09.009} {\bibfield  {journal} {\bibinfo
  {journal} {Comput.Phys.Commun.}\ }\textbf {\bibinfo {volume} {184}},\
  \bibinfo {pages} {222} (\bibinfo {year} {2013})}\BibitemShut {NoStop}%
\end{thebibliography}%

\end{document}